\documentclass[12pt]{iopart}
\usepackage{graphicx}

\begin{document}

\title{Spin waves in a complex magnetic system: nonextensive approach}

\author{D O Soares-Pinto$^{1,2}$, M S Reis$^2$, R S Sarthour$^1$ and I S Oliveira$^1$}

\address{$^1$ Centro Brasileiro de Pesquisas
F\'{\i}sicas, Rua Dr. Xavier Sigaud 150, 22290-180, Rio de
Janeiro, Brazil}
\address{$^2$ Departamento de F\'{\i}sica and CICECO, Universidade de Aveiro, 3810-193,
Aveiro, Portugal}

\ead{\mailto{dosp@cbpf.br},\mailto{marior@ua.pt},\mailto{sarthour@cbpf.br},\mailto{ivan@cbpf.br}}

\begin{abstract}
In this paper we analyze the spin-wave excitations (magnons) of an
inhomogeneous spin system within the Boltzmann-Gibbs framework and
then connect the results with the nonextensive approach (in the
sense of Tsallis statistics). Considering an equivalence between
those two frameworks, we could connect the entropic parameter $q$
with moments of the distribution of exchange integrals of the
inhomogenous system. It ratifies the idea that the entropic
parameter is connected to the microscopic properties of the
system.
\end{abstract}


\section{Introduction}

Inspired in multifractals, Tsallis proposed a generalization of
Boltzmann-Gibbs entropy ($S_{BG}$)~\cite{1988_JSP_52_479}
\begin{equation}\label{eq.01}
\fl S_{q} = k \frac{1-\sum_{i} p_{i}^{q}}{q-1}\qquad (q\,\in\,
\Re)
\end{equation}
where $q$ is the entropic parameter for a specific system and is
connected to its dynamics as recently
proposed~\cite{2006_PhysRevB_73_092401, 2006_EPJB_50_99}, $p_{i}$
are the probabilities satisfying $\sum_{i} p_{i}=1$, $k$ is a
constant, and $\lim_{q\rightarrow 1}S_{q} = S_{BG}$. This entropy
for a system composed of two independent parts $A$ and $B$, such
as the probability is given by $P(A\bigcap B) = P(A)\,P(B)$, has
the interesting property of nonextensivity (see for example
Refs.~\cite{1999_BrJP_29_1, 2004_CMT_16_223})
\begin{equation}
S_{q}(A+B) = S_{q}(A)+S_{q}(B)+(1-q)S_{q}(A)S_{q}(B)
\end{equation}
Besides representing a generalization, $S_{q}$, like $S_{BG}$, is
nonnegative, concave, Lesche-stable ($\forall\, q > 0$). Recently
it has been shown that it is also {\it extensive} for some sorts
of correlated systems~\cite{2005_PNAS_102_15377}.

Tsallis statistics, or nonextensive statistics, attempt to handle
some anomalies that appear in physical problems which cannot be
treated with Boltzmann-Gibbs ($BG$) statistics, for instance,
long-range correlations, intrinsic cooperativity, multifractal
structure, dissipation in mesoscopic scale, strong non-Markovian
microscopic memory, etc~\cite{2002_CSF_13371}. These anomalies
have the common characteristic of presenting power-laws, instead
of the ordinary exponential-laws, which is also a characteristic
of some complex systems. Its applicability ranges from protein
folding~\cite{2006_JCC_27_1142} to financial
markets~\cite{2007_EPJB_55_161}, and from
turbulence~\cite{1996_PhysRevE_53_4754} to cosmic
rays~\cite{2004_PhysA_331_173}. For example, in condensed matter
problems we can cite Ising ferromagnets, Landau diamagnetism,
electron-phonon systems, tight-binding-like Hamiltonians, metallic
and superconductor systems~\cite{website_TEMUCO}. In adittion, an
interesting example is in~\cite{2002_PhysRevB_66_4417,
2003_PhysRevB_68_014404} where the authors could predict some
peculiar magnetic properties of manganites using nonextensive
statistics like nanoscale inhomogeneity and phase coexistence,
fractal structures, and long-range
interactions~\cite{1999_Science_283_2034, 2001_PhysRep_344_1,
2005_Sci_309_257}.

Herein, we present some results comparing an inhomogeneous spin
system within $BG$ framework and a homogeneous spin system into a
nonextensive approach. This comparison lead us to a connection
between the nonextensive parameter $q$ with specific moments of
the distribution of the exchange integral of the inhomogeneous
system. Thus, the spin waves in a inhomogeneous magnetic media can
be described by the nonextensive statistics and the entropic
parameter is connected to the microscopic properties of the
system, as previously shown for other systems by Beck
\cite{2001_PhysRevLett_87_180601}, Beck and Cohen
\cite{2003_PhysA_322_267}, Wilk and W{\l}odarczyk
\cite{2000_PhysRevLett_84_2770}, Reis et
al.~\cite{2006_PhysRevB_73_092401}, and therefore can be seen as a
measurement of its complexity.

\section{Spin-waves}

\subsection{Magnons within inhomogeneous medium: Boltzmann-Gibbs
framework}

In a ferromagnet at $T=0$K all the spins have the maximum
projection $S$ along the $z$ direction; this is the ground state
configuration~\cite{livro_ivan, livro_apg}. Letting the spin
system be in thermal contact with a reservoir, as the temperature
increases, it will leave its ground state, the projections along
the quantization direction will be reduced, and a wave-like
perturbation will flow through the spin system; that is the
spin-wave (magnons). The spin-wave theory leads to the description
of the magnetism of ferromagnets at low temperatures, in the
regime where the total angular momentum is close to its the
projection along the $z$ direction, $S\cong S^{z}$.

We will consider a system of $N$ spins, each one interacting with
$z$ neighbors in a inhomogeneous way, and in the presence of a
magnetic field $B_{0}$. Thus, the Hamiltonian for this
inhomogeneous magnetic system is given by~\cite{livro_ivan,
livro_majlis, livro_mattis}
\begin{equation}\label{eq.02}
\fl\mathcal{H} = - \sum_{\bf R,d}J_{d}({\bf R})\,{\bf S}_{\bf
R}\cdot{\bf S}_{\bf R+d}-h\sum_{\bf R}S_{{\bf R}z}
\end{equation}
in which $J_{d}({\bf R})>0$ (always ferromagnetic) describes the
inhomogeneity of the media, i.e., there is a distribution of
exchange interactions $f(J)$, and $h=g\,\mu_{B}\,B_{0}$ plays the
role of the applied magnetic field. The whole Hamiltonian can be
re-written in terms of the collective motion operators in order to
give us the magnetization per unit of volume. One can write the
spin operators as
\begin{equation}\label{eq.03}
\fl{\bf S}_{\bf  R}\cdot{\bf S}_{\bf R+d} =
\frac{1}{2}\,\left[S_{\bf R}^{+}S_{\bf R+d}^{-}+ S_{\bf
R}^{-}S_{\bf R+d}^{+}\right]+ S_{{\bf R}z}S_{{\bf R+d}z}.
\end{equation}

The Holstein-Primakoff transformation of spin operators, at low
temperatures, is given by
\begin{equation}\label{eq.04}
\fl S_{\bf R}^{+}\approx\sqrt{2\,S}\,a_{\bf R}\,\mbox{ and
}\,S_{\bf R}^{-}\approx\sqrt{2\,S}\,a_{\bf R}^{+}
\end{equation}
in which $a_{\bf R}^{+}$ and $a_{\bf R}$ obey the commutation
relation $\left[a_{\bf R},a_{\bf R}^{+}\right] = 1$. The operators
$a_{\bf R}$ and $a_{\bf R}^{+}$ can be written in terms of the
collective motion of the system
\begin{equation}\label{eq.05}
\fl a_{\bf R} = \frac{1}{\sqrt{N}}\sum_{k} e^{i{\bf k}\cdot{\bf
R}}\,b_{k}\,\mbox{ and }\,a_{\bf R}^{+} =
\frac{1}{\sqrt{N}}\sum_{k} e^{-i{\bf k}\cdot{\bf R}}\,b_{k}^{+}
\end{equation}
in which $[b_{k},b_{k}^{+}]=1$.

Thus, one can rewrite the exchange term of the Hamiltonian (\ref{eq.02})
\begin{equation}\label{eq.06}
\fl -\sum_{\bf R,d}J_{d}({\bf R})\,{\bf S}_{\bf R}\cdot{\bf
S}_{\bf R+d}= - \frac{1}{2}\,\sum_{\bf R,d}J_{d}({\bf
R})\,\left[S_{\bf R}^{+}S_{\bf R+d}^{-}+ S_{\bf R}^{-}S_{\bf
R+d}^{+}\right]-\sum_{\bf R,d}J_{d}({\bf R})\,S_{{\bf R}z}S_{{\bf
(R+d)}z}.
\end{equation}

Using the low-temperature Holstein-Primakoff transformation, in
terms of the collective motion operators, we have

\begin{equation}\label{eq.07}
\fl\frac{1}{2}\,\sum_{\bf R,\,d}J_{d}({\bf R})\left[S_{\bf
R}^{+}S_{\bf R+d}^{-}+ S_{\bf R}^{-}S_{\bf R+d}^{+}\right]=
\sum_{k}\,\left(\frac{2\,S}{N}\sum_{\bf R,\,d}J_{d}({\bf
R})\cos({\bf k}\cdot{\bf d})\right)\,b_{k}^{+}\,b_{k}
\end{equation}
and
\begin{equation}\label{eq.08}
\fl\sum_{\bf R,\,d}J_{d}({\bf R})\,S_{{\bf R}z}S_{{\bf R+d}z}=
N\,S^{2}\left(\frac{1}{N}\sum_{\bf R,\,d}J_{d}({\bf R})\right)-
2\,S\left(\frac{1}{N}\sum_{\bf R,\,d}J_{d}({\bf
R})\right)\sum_{k}b_{k}^{+}\,b_{k}.
\end{equation}
Note that we have excluded the magnon-magnon interaction,
represented by the term of $n_{k}\,n_{k}$.

In terms of these operators,
\begin{equation}\label{eq.09}
\fl S_{{\bf R}z} = S - a_{\bf R}^{+}a_{\bf R}
\end{equation}
where $S$ is the spin value per site and therefore the second term
of the Hamiltonian can be written as

\begin{equation}\label{eq.10}
\fl -h\sum_{\bf R}S_{{\bf R}z}=
h\sum_{k}\,b_{k}^{+}\,b_{k}-h\,N\,S
\end{equation}
where $N$ is the number of sites. Hence, the Hamiltonian becomes
\begin{equation}\label{eq.11}
\fl\mathcal{H} = -(h\,N\,S + N\,S^{2}\,J) + \sum_{k}\,\left(h +
2\,S\,J - \frac{2\,S}{N}\sum_{\bf R,\,d}J_{d}({\bf R})\cos({\bf
k}\cdot{\bf d})\right)\,n_{k}.
\end{equation}
where $n_{k} = b_{k}^{+}\,b_{k}$ is the Boson number operator and
$J \equiv \frac{1}{N}\sum_{\bf R,\,d}J_{d}({\bf R})$. The first
two terms represent the fundamental state of the system, or the
total energy without excitations. The term that describes the
magnons is the second one. It has the form
$\sum_{k}\hbar\,\omega_{k}\,n_{k}$ and gives the dispersion
relation for this inhomogeneous magnetic system
\begin{equation}\label{eq.12}
\fl\hbar\,\omega_{k}= h + 2\,S\,J - \frac{2\,S}{N}\sum_{\bf
R,\,d}J_{d}({\bf R})\cos({\bf k}\cdot{\bf d}).
\end{equation}

For large wave length, one may write

\begin{equation}\label{eq.13}
\fl\sum_{\bf R,\,d}J_{d}({\bf R})\cos({\bf k}\cdot{\bf d}) \approx
\sum_{\bf R,\,d}J_{d}({\bf R})\left[1-\frac{1}{2}({\bf k}\cdot{\bf
d})^{2}\right]
\end{equation}
and $({\bf k}\cdot{\bf d})^{2} = k^{2}\,a^{2}$, being $a$ the
lattice parameter. Thus, the dispersion relation is

\begin{equation}\label{eq.14}
\fl\hbar\,\omega_{k} \approx h+ k^{2}\,\mathcal{D}(J).
\end{equation}
where $\mathcal{D}(J)=a^{2}\,z\,S\,J$ is the stiffness parameter
and $z$ is the number of first neighbors.

Since the interaction varies between spins, one may consider that
it has a distribution $f(J)$. Thus, the average magnetization
varies with respect to the saturation value of the magnetization
and its variation is given by
\begin{equation}\label{eq.15}
\fl\langle\Delta\,m\rangle =
\frac{g\,\mu_{B}}{2\,\pi^{2}}\int_{0}^{\infty}dJ\,f(J)
\int_{0}^{\infty}dk\,k^{2}\langle n_{k}\rangle_{J}
\end{equation}
where $\langle n_{k}\rangle_{J}$ is the Planck distribution. Then,
(\ref{eq.15}) becomes
\begin{eqnarray}\label{eq.16.17}
\fl\langle\Delta\,m\rangle &=&
\frac{g\,\mu_{B}}{2\,\pi^{2}}\int_{0}^{\infty}dJ\,f(J)
\int_{0}^{\infty}dk\,\frac{k^{2}}{\e^{(k^{2}\,\mathcal{D}(J)+h)/k_{B}\,T}-1}\\
\fl &=& \frac{g\,\mu_{B}}{4\,\pi^{2}}\int_{0}^{\infty}dJ\,f(J)
\frac{k_{B}\,T}{\mathcal{D}(J)^{3/2}}\int_{h/k_{B}\,T}^{\infty}dx\,\frac{(k_{B}\,T\,x-h)^{1/2}}{\e^{x}-1}
\end{eqnarray}
where $x = (k^{2}\,\mathcal{D}(J)+h)/k_{B}\,T$.

For $B_{0} = 0$, i.e., $h = 0$, the inner integral becomes
\begin{equation}\label{eq.18}
\fl\int_{0}^{\infty}\frac{x^{1/2}\,dx}{\e^{x}-1} =
\frac{\sqrt{\pi}}{2}\zeta(3/2)
\end{equation}
where $\zeta(n)$ is the Riemann Zeta function. The volume
magnetization variation due to the magnon excitation of a
inhomogeneous is then given by
\begin{equation}\label{eq.19}
\fl\langle\Delta\,m\rangle = \frac{\zeta(3/2)\,g\,
\mu_{B}}{8\,\pi^{3/2}}\left(\frac{k_{B}\,T}{a^{2}\,z\,S}\right)^{3/2}
\int_{0}^{\infty}dJ\,\frac{f(J)}{J^{3/2}}=\frac{\zeta(3/2)\,g\,
\mu_{B}}{8\,\pi^{3/2}}\left(\frac{k_{B}\,T}{a^{2}\,z\,S}\right)^{3/2}\langle
J^{-3/2}\,\rangle.
\end{equation}

It is important to emphasize that the volume magnetization change
of the inhomogeneous system has a $T^{3/2}$ dependence (like the
homogenous case) and also depends on the $-3/2$ moment of the
distribution of exchange integrals $\langle J^{-3/2}\rangle$. This
exponent is expected since 3 is related to the dimension of the
system and 2 refer to the dynamics, i.e., came from the dispersion
relation (\ref{eq.14}).

\subsection{Magnons within homogeneous medium: Nonextensive
framework}

The dynamics of a system is given by its Hamiltonian $\mathcal{H}$
and the wave number $k$, defined by $\mathcal{H}$, is,
consequently, related to the dynamics. On the other hand, the
statistics of a system is given by an average over a great number
of variables; and lies, for instance, in the number of Bosons
$n_{k}$ for each wave number $k$. This average over weighted
states makes it possible to obtain the relation of microscopic
physical properties and macroscopic thermodynamic quantities such
as the volume magnetization variation. The nonextensive approach
proposes a change on the statistics of the system, not on its
dynamics. Thus, we assume an equivalent Hamiltonian as
(\ref{eq.02}), but homogeneous in this framework, i.e., the
exchange integral can be taken out of the sum. The dispersion
relation is therefore given by $\epsilon_{k} =
a^{2}\,S\,\mathcal{J}\,k^{2}$, where $\mathcal{J}$ is the exchange
integral of this homogeneous system. The volume magnetization in
this nonextensive scenario can be written
as~\cite{livro_souzatsallis}
\begin{equation}\label{eq.20}
\fl\langle\Delta\,m\rangle_{q} =
\frac{g\,\mu_{B}}{2\,\pi}\,\int_{0}^{\infty}
dk\,k^2\,\langle\,n_{k}\rangle_{q,\mathcal{J}}
\end{equation}
in which $n_{k} = b_{k}^{+}\,b_{k}$ is the Boson number operator
and $\langle ...\rangle_{q}$ is not the standard Planck
distribution, but its $q$-version, i.e., generalized Plank
distribution
\begin{equation}\label{eq.21}
\fl\langle\,n_{k}\rangle_{q,\mathcal{J}} =
\frac{\Tr\left\{n_{k}\,\rho^q\right\}}{\Tr\left\{\rho^q\right\}}
=\frac{\sum_{n_{k}=0}^{\infty}n_{k}\,[1-(1-q)(\beta\,n_{k}\,\epsilon_{k})]^{\frac{q}{1-q}}}
{\sum_{n_{k}=0}^{\infty}[1-(1-q)(\beta\,n_{k}\,\epsilon_{k})]^{\frac{q}{1-q}}}.
\end{equation}

Using the dispersion relation described above and making
(\ref{eq.20}) dimensionless, one gets
\begin{equation}\label{eq.22}
\fl\langle\Delta\,m\rangle_{q}
=\frac{g\,\mu_{B}}{4\,\pi^{2}}\,\left(\frac{k_{B}\,T}{a^{2}\,S\,\mathcal{J}}\right)^{3/2}\,
\int_{0}^{\infty}\,dx\,x^{1/2}\,f(x,q)
\end{equation}
where
\begin{equation}\label{eq.23}
\fl f(x,q) =
\frac{\sum_{n_{k}=0}^{\infty}n_{k}\,[1-(1-q)(x\,n_{k})]^{\frac{q}{1-q}}}
{\sum_{n_{k}=0}^{\infty}[1-(1-q)(x\,n_{k})]^{\frac{q}{1-q}}}.
\end{equation}

Finally, the magnetization can be written as
\begin{equation}\label{eq.24}
\fl\langle\Delta\,m\rangle_{q}
=\frac{g\,\mu_{B}}{4\,\pi^{2}}\,\left(\frac{k_{B}\,T}{a^{2}\,S\,\mathcal{J}}\right)^{3/2}\,F(q)
\end{equation}
where $F(q)$ is the integral which appears in Eq.(\ref{eq.22}).
One can see that the magnetization in this scenario has the same
$T^{3/2}$ behavior as in (\ref{eq.19}). It is a consequence that
neither the dynamics ($\epsilon_{k}\propto k^{2}$) nor the
dimension ($d = 3$) have being changed. All the information about
the homogeneity and/or inhomogeneity of the system is into the
statistics and, consequently, into the coefficient of the
magnetization change.

An analytical connection between  the entropic parameter $q$ and
the volume magnetization variation can be obtained at the limit
$(q-1)\rightarrow 0$. At this limit, we can write (\ref{eq.22}) as
\begin{equation}\label{eq.25}
\fl\langle n_{k}\rangle_{q,\mathcal{J}} =
\frac{\frac{1}{e^{q\,\beta\,\epsilon_{k}}-1}+\frac{1}{2}\,
(\beta\,\epsilon_{k})^{2}\,(q-1)\,\frac{1+4\,e^{q\,\beta\,\epsilon_{k}}+e^{2\,q\,\beta\,\epsilon_{k}}}
{(e^{q\,\beta\,\epsilon_{k}}-1)^3}}{1+\frac{1}{2}\,(\beta\,\epsilon_{k})^{2}\,(q-1)\,\frac{e^{q\,\beta\,\epsilon_{k}}+1}
{(e^{q\,\beta\,\epsilon_{k}}-1)^2}}
\end{equation}

Thus the volume magnetization change is now given by
\begin{equation}\label{eq.26}
\fl\langle\Delta\,m\rangle_{q} =
\frac{g\,\mu_{B}}{2\,\pi^{2}}\,\int_{0}^{\infty}\,dk\,k^{2}\langle
n_{k}\rangle_{q,\mathcal{J}}
=\frac{g\,\mu_{B}}{4\,\pi^{2}}\left(\frac{k_{B}\,T}{a^{2}\,z\,S\,\mathcal{J}}\right)^{3/2}
\left[\Gamma_{q}^{'}+\frac{(q-1)}{2}\Gamma_{q}^{''}\right]
\end{equation}
in which $\Gamma_{q}^{'}$ and $\Gamma_{q}^{''}$ are dimensionless
integrals
\begin{equation}\label{eq.27}
\fl\Gamma_{q}^{'} = \int_{0}^{\infty}
\frac{x^{1/2}\,dx}{\left[e^{q\,x}-1\right]\,
\left[1+\frac{(q-1)}{2}\,x^2\,\frac{\left[e^{q\,x}+1\right]}{\left[e^{q\,x}-1\right]^2}\right]}
= \frac{\sqrt{\pi}}{2}\,\zeta\left(\frac{3}{2}\right)-
5.2277\,(q-1)
\end{equation}
and
\begin{equation}\label{eq.28}
\fl\Gamma_{q}^{''} = \int_{0}^{\infty}
\frac{\left[1+4\,e^{q\,x}+e^{2\,q\,x}\right]
\,x^{5/2}\,dx}{\left[e^{q\,x}-1\right]^3\,\left[1+\frac{(q-1)}{2}\,x^2\,\frac{\left[e^{q\,x}+1\right]}
{\left[e^{q\,x}-1\right]^2}\right]} = 16.4154 - 68.6515\,(q-1).
\end{equation}

As it is an approximation for $q$ close to 1, there is no
necessity of terms higher than $(q-1)$, so the second term of
$\Gamma_{q}^{''}$ can be neglected. Thus,
\begin{equation}\label{eq.29}
\fl\langle\Delta\,m\rangle_{q}=\frac{g\,\mu_{B}}{4\,\pi^{2}}\left(\frac{k_{B}\,T}{a^{2}\,z\,S\,\mathcal{J}}\right)^{3/2}
\left[\frac{\sqrt{\pi}}{2}\,\zeta\left(\frac{3}{2}\right)+2.98\,(q-1)\right].
\end{equation}

One can see that, in this approximation, the magnetization is
directly related to the entropic index and when $q\rightarrow 1$,
it recovers the usual result (homogeneous case within BG
statistics).

\section{Mean field approximation and the critical temperature}

Let us consider that the two systems discussed before have the
same critical temperature, as already done in the
literature~\cite{2003_PhysRevB_68_014404}. Considering the
Hamiltonian (\ref{eq.02}) for an inhomogeneous magnetic system
within the mean field approximation, one may change the quantum
operator ${\bf S}_{{\bf R+d}}$ for its thermal average $\langle
{\bf S}_{{\bf R+d}}\rangle_{T}$. Thus, considering $z$ first
neighbors of an atom on the $R^{th}$ site of the lattice, its
Hamiltonian becomes
\begin{equation}\label{eq.30}
\fl\mathcal{H}_{tr}^{R}=- \langle J\rangle \,z\,\mathbf{S}_{{\bf
R}}\cdot \langle\mathbf{S}_{{\bf R+d}}\rangle_{T}
\end{equation}
where in this approximation we can consider the exchange
interaction between the spins as an average value $\langle
J\rangle$. It is reasonable, because all of the spins, in the mean
field approximation, interacts with all others spins in the same
way.

For the above Hamiltonian it is straightforward to obtain the
critical temperature
\begin{equation}\label{eq.31}
\fl T_{c}= \frac{z\,S\,(S+1)}{3\,k_{B}}\,\langle J\rangle.
\end{equation}

An analogous calculation can be done in the nonextensive scenario
\cite{2002_EurophysLett_58_42}. The generalized Brillouin
function~\cite{2006_PhysRevB_73_092401} gives us the critical
temperature
\begin{equation}\label{eq.32}
\fl T_{c}^{(q)}= \frac{z\,S\,(S+1)}{3\,k_{B}}\,q\,\mathcal{J}
\end{equation}
in which $\mathcal{J}$ is the exchange integral in this framework.
The relation between these two temperatures is given by
\cite{2003_PhysRevB_68_014404, 2002_EurophysLett_58_42}
\begin{equation}\label{eq.33}
\fl T_{c}^{(q)} = q\,T_{c}^{(1)}.
\end{equation}

Thus, using (\ref{eq.31}), (\ref{eq.32}) and (\ref{eq.33}) one
find the relation between the two exchange integrals
\begin{equation}\label{eq.34}
\fl \mathcal{J} = \langle J\rangle.
\end{equation}
that is, the exchange integral in nonextensive framework is
equivalent to and average of the inhomogeneous one. This result is
expected since, as already discussed above, we are not changing
the dynamics of the system, but only but only the statistical
treatment, which is used to calculate its thermodynamical
properties.

\section{Equivalence between the two frameworks}

Comparing the magnetization change per unit of volume in the
inhomogeneous framework (\ref{eq.20}) with its analogous in the
nonextensive scenario (\ref{eq.24}), one finds that
\begin{equation}\label{eq.35}
\fl F(q) =
\frac{\sqrt{\pi}}{2}\,\zeta\left(\frac{3}{2}\right)\frac{\langle
J^{-3/2}\,\rangle}{\mathcal{J}^{-3/2}} =
\frac{\sqrt{\pi}}{2}\,\zeta\left(\frac{3}{2}\right)\frac{\langle
J^{-3/2}\,\rangle}{\langle J\rangle^{-3/2}}
\end{equation}

\begin{figure}[tbh]
\begin{center}
\includegraphics[width=0.5\columnwidth,angle=0]{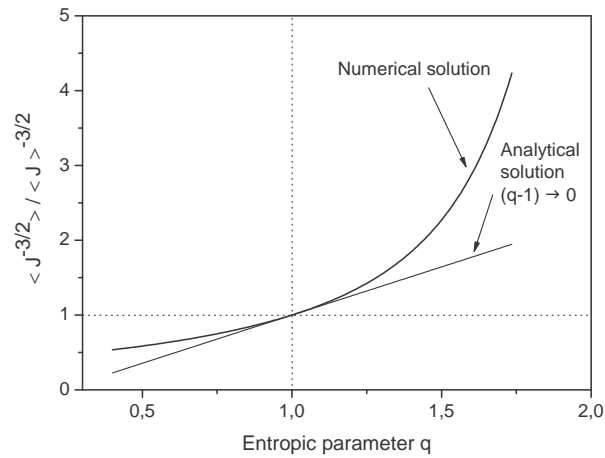}
\end{center}
\caption{The entropic parameter $q$ is connected to specific
moments of the distribution of exchange integrals. This result is
valid for any $f(J)$ and shows that the entropic parameter is
connected to the physical properties of the system.}\label{figura}
\end{figure}

The above equation is a relation of the $q$ parameter and moments
of the exchange interaction distribution $f(J)$. Figure
(\ref{figura}) presents the expression above numerically solved
for $q\in [0.1,1.9]$. This procedure comparing the magnetization
was already been done~\cite{2003_PhysRevB_68_014404}, where those
authors have found similar results inspired in
Superstatistics~\cite{2003_PhysA_322_267}.

\newpage
An analytical connection between the entropic parameter $q$ and
the specific moments of the exchange integral of the inhomogeneous
magnetic media can be obtained using the expression for the volume
magnetization change in the limit $(q-1)\rightarrow 0$. Comparing
the (\ref{eq.20}) and  (\ref{eq.27}) one gets
\begin{equation}\label{eq.36}
\fl \frac{\sqrt{\pi}}{2}\,\zeta\left(\frac{3}{2}\right)+
2.98\,(q-1) =
\frac{\sqrt{\pi}}{2}\,\zeta\left(\frac{3}{2}\right)\frac{\langle
J^{-3/2}\,\rangle}{\langle J\rangle^{-3/2}}
\end{equation}
or
\begin{equation}\label{eq.37}
\fl (q-1) = 0.78\,\left[\frac{\langle J^{-3/2}\,\rangle}{\langle
J\rangle^{-3/2}}-1\right].
\end{equation}

The result above is also valid for any $f(J)$ and shows that the
entropic parameter is connected to the physical properties of the
system~\cite{2006_PhysRevB_73_092401, 2001_PhysRevLett_87_180601,
2003_PhysA_322_267, 2000_PhysRevLett_84_2770}.

\section{Final remarks}

Summarizing, in the present work we have shown that the $q$
parameter can be seen as a measurement of the inhomogeneity of a
magnetic system. It ratifies previous
works~\cite{2006_PhysRevB_73_092401, 2006_EPJB_50_99} in which,
inspired in Superstatistics~\cite{2003_PhysA_322_267}, the
entropic parameter $q$ was related to the first and second moments
of the distribution of magnetic moments of manganites
\begin{equation}\label{eq.38}
\fl q\,(2-q)^{2} = \frac{\langle\mu^{2}\rangle}{\langle\mu\rangle^{2}}
\end{equation}
and was also experimentally verified. Thus, the present work
corroborates the idea that changing the usual Boltzmann$-$Gibbs
statistics to one that is able to describe power-laws (Tsallis
statistics), one can characterize systems that has special
features like inhomogeneities; Nonextensivity is therefore a key
role to describe complex systems.

\ack DOSP would like to thanks the brazilian funding agency CAPES
for the financial support at Universidade de Aveiro at Portugal.
This work was partially supported by the brazilian funding
agencies CNPq and CAPES.

\end{document}